\documentclass[12pt]{article}
\usepackage{graphics,epsfig}
\usepackage[english]{babel}
\newcommand{\cinst}[2]{$^{\mathrm{#1}}$~#2\par}
\newcommand{\crefi}[1]{$^{\mathrm{#1}}$}

\begin{document}

%\begin{titlepage}

%%%%%%%%%%%%%%%%%%%%%%%% COVER PAGE
\begingroup

\vglue.5cm
%\end{minipage}

%&
%\begin{minipage}[t]{7cm}
%\hspace{11cm} {\bf DRAFT 1}

%\end{minipage}
%\end{tabular*}
%}
\vspace{1.3cm}
\begin{center}

{\Large{\bf The A - dependence of $K^{0}$ and $\Lambda$\\
\vspace{0.2cm} neutrinoproduction on nuclei}}
\end{center}

\vspace{1.cm}

\begin{center}
{\large SKAT Collaboration}

 N.M.~Agababyan\crefi{1}, V.V.~Ammosov\crefi{2},
 M.~Atayan\crefi{3},\\
 N.~Grigoryan\crefi{3}, H.~Gulkanyan\crefi{3},
 A.A.~Ivanilov\crefi{2},\\ Zh.~Karamyan\crefi{3},
V.A.~Korotkov\crefi{2}

\setlength{\parskip}{0mm}
\small
%\HRule\\

\vspace{1.cm} \cinst{1}{Joint Institute for Nuclear Research,
%Dubna, Russia}\\ \cinst{2}{Institute for High Energy Physics,
%Protvino, Russia} \cinst{3}{Yerevan Physics Institute, Armenia}\\
Dubna, Russia} \cinst{2}{Institute for High Energy Physics,
Protvino, Russia} \cinst{3}{Yerevan Physics Institute, Armenia}
\end{center}
\vspace{100mm}

{\centerline{\bf YEREVAN  2005}}

%\end{titlepage}
\newpage
\begin{abstract}

For the first time, the A- dependence of the production of $K^0$,
$\Lambda$ and, for comparison, $\pi^-$ mesons is investigated in
neutrinonuclear reactions, using the data obtained with SKAT
bubble chamber. An exponential parametrization ($\sim A^{\beta}$)
of the particle yields results in ${\beta}_{V^0} = 0.20 \pm 0.05$
for $V^0$ particles (combined $K^0$ and $\Lambda$), while for
$\pi^-$ mesons the A- dependence is much weaker, ${\beta}_{\pi^-}
= 0.068 \pm 0.007$. A nuclear enhancement of the ratio $K^0/\pi^-$
is found; this ratio increases from $0.055 \pm 0.013$ for $\nu N$-
interactions up to $0.070 \pm 0.011$ at $A \approx 21$ and $0.099
\pm 0.011$ at $A \approx 45$. It is observed, that the
multiplicity rise of $V^0$'s occures predominantely in the
backward hemisphere of the hadronic c.m.s. It is shown, that the
A- dependence of the nuclear enhancement of the ${\Lambda}^0$ and
$\pi^-$ yields can be reproduced in the framework of a model,
incorporating the secondary intranuclear interactions of pions
originating from the primary $\nu N$- interactions, while only
(29$\pm$9)\% of that for $K^0$ at $A \approx 45$ can be attributed
to intranuclear interactions.

\end{abstract}

%\end{titlepage}

\newpage
\section{Introduction}

%\vglue2cm
The leptoproduction processes on nuclei provide a valuable
information on the space-time structure of the quark string
fragmentation and the hadron formation. Hitherto the detailed
experimental data are available for the pion leptoproduction
processes. The data for strange particles are comparatively
scarce. In particular, no data are available on the A- dependence
of their leptoproduction. Meanwhile, this process can carry an
additional information concerning the nuclear medium influence on
the dynamics of the quark string fragmentation, because the yield
of strange particles is rather sensitive (unlike pions) to the
string tension which can, in principle, be affected by the nuclear
medium.\\ This work is aimed to infer the first experimental data
on the A- dependence of yields of neutral strange particles ($K^0$
mesons and $\Lambda$ hyperons) in neutrinonuclear interactions. In
Section 2, the experimental procedure is briefly described. The
experimental data on the A- dependences of the mean multiplicities
and inclusive spectra of $K^0$ and $\Lambda$ are presented in
Section 3 and discussed in Section 4, where a comparison of the
data with the model predictions is presented. The results are
summarized in Section 5.

\section{Experimental procedure}

The experiment was performed with SKAT bubble chamber \cite{ref1},
exposed to a wideband neutrino beam obtained with a 70 GeV primary
protons from the Serpukhov accelerator. The chamber was filled
with a propan-freon mixture containing 87 vol\% propane ($C_3H_8$)
and 13 vol\% freon ($CF_3Br$) with the percentage of nuclei
H:C:F:Br = 67.9:26.8:4.0:1.3 \%. A 20 kG uniform magnetic field
was provided within the operating chamber volume. The selection
criteria of the properly reconstructed charged current
interactions and the procedure of the reconstruction of the
neutrino energy $E_{\nu}$ can be found in our previous
publications (\cite{ref2,ref3} and references therein). Each event
was given a weight (depending on the charged particle
multiplicity) which corrects for the fraction of events excluded
due to improperly reconstruction. The events with 3 $<E_{\nu}<$ 30
GeV were accepted, provided that the invariant mass $W$ of the
hadronic system exceeds  1.5 GeV. No restriction was imposed on
the transfer momentum squared $Q^2$. The number of accepted events
was 5987 (6953 weighted events). The mean values of the
kinematical variables are: $<E_{\nu}>$ = 9.8 GeV, $<W>$ = 2.7 GeV,
$<W^2>$ = 7.9 GeV$^2$, $<Q^2>$ = 2.5 (GeV/$c)^2$, and the mean
energy transferred to the hadronic system $<{\nu}>$ = 5.1 GeV.\\
The selection criteria for the decay of neutral strange particles
and the procedure of their identification were similar to those
applied in \cite{ref2}. The number of the accepted neutral strange
particles  ($V^{\circ}$'s) was 110 out of which 46(64) had the
biggest probability to be identified as $K^{\circ}({\Lambda})$.
The corresponding average multiplicities, corrected for the decay
losses, are $<n_{V^{\circ}}>$ = (7.11$\pm$0.68)$\cdot 10^{-2}$,
$<n_{K^{\circ}}>$ = (4.29$\pm$0.63)$\cdot 10^{-2}$,
$<n_{\Lambda}>$ = (2.82$\pm$0.35)$\cdot 10^{-2}$.\\ For the
further analysis the whole event sample was subdivided, using
several topological and kinematical criteria \cite{ref4}, into
three subsamples: the 'cascade' subsample $B_S$ with a sign of
intranuclear secondary interactions, the 'quasiproton' $(B_p)$ and
'quasineutron' $(B_n)$ subsamples for which no sign of secondary
interactions was observed. The corresponding weighted event
numbers for $B_S$, $B_p$ and $B_n$ subsamples are 3654, 1653 and
1646, respectively. About 40\% of subsample $B_p$ is contributed
by interactions with free hydrogen. Weighting the 'quasiproton'
events with a factor of 0.6, one can compose a 'quasinucleon'
subsample $B_N = B_n + 0.6 B_p$ and a 'pure' nuclear subsample
$B_A = B_S + B_N$ having an effective atomic weight
$A_{eff}\approx$ 21 (the definition of $A_{eff}$ for the composite
target will be done below in Section 4). In the next sections, we
use the data obtained for subsamples $B_N$ and $B_A$, as well as
the published data \cite{ref2} on neutrino-freon interactions
(with $A_{eff} \approx$ 45), in order to extract the A- dependence
of the yields of the neutral strange particles and to infer an
information about the nuclear medium influence on their
production. We have verified, that the mean values of the
kinematical variables (quoted above) are consistent in the all
three data sets (at $A_{eff}$ = 1, 22 and 45).

\section{The A- dependence of the mean multiplicities and
inclusive spectra}

The mean multiplicities $<n_{K^{\circ}}>$, $<n_{\Lambda}>$ and,
for comparison, $<n_{{\pi}^-}>$ for ${\pi}^-$ mesons, as well as
the ratio $R(K^{\circ}/\pi^-)$ = $<n_{K^{\circ}}>/<n_{\pi^-}>$ are
presented in Table 1 for the 'quasinucleon' and nuclear subsamples
and for neutrino-freon interactions ($A_{eff} \approx 45$)
\cite{ref2}.

\begin{table}[ht]
\begin{center}
\begin
{tabular}{|l c c c c|}
  \hline
%\multicolumn{5}{|c|}{} \\
% &\multicolumn{4}{c|}{4 $< W^2 <$ 25
%GeV$^2$}\\  \multicolumn{5}{|c|}{} \\ $h^+(x_F > 0)$
$A_{eff}$&$<n_{K^{\circ}}>$&$<n_{\Lambda}>$&$<n_{\pi^-}>$&$R(K^{\circ}/\pi^-)$
\\ \hline 1&0.030$\pm$0.007&0.018$\pm$0.004&0.55$\pm$0.01&0.055$\pm$0.013 \\
21&0.044$\pm$0.006&0.030$\pm$0.004&0.63$\pm$0.01&0.070$\pm$0.011
\\
45&0.071$\pm$0.008&0.031$\pm$0.004&0.72$\pm$0.01&0.099$\pm$0.011
\\ \hline
\end{tabular}

\end{center}
\caption{The mean multiplicities $<n_{K^{\circ}}>$,
$<n_{\Lambda}>$, $<n_{\pi^-}>$ and the ratio $R(K^{\circ}/\pi^-)$
at different $A_{eff}$.}
\end{table}

\noindent It should be noted, that the quoted values of
$<n_{K^{\circ}}>_N$ = 0.030$\pm$0.007 and $<n_{\Lambda}>_N$ =
0.018$\pm$0.004 for the 'quasinucleon' subsample do not contradict
the available data around $E_{\nu}\sim$ 10 GeV obtained for $\nu$p
interactions \cite{ref5,ref6}. The A- dependence of the data
presented in Table 1 is approximated as $\sim A^{\beta}_{eff}$
(Fig. 1). The fitted values of the slope parameter $\beta$ are:
${\beta}_{K^0} = 0.225 \pm 0.070$, ${\beta}_{\Lambda} = 0.147 \pm
0.069$ and ${\beta}_{{\pi}^-} = 0.068 \pm 0.007$. Similarly, for
the combined data on the neutral strange particles ($V^0 \equiv
K^0 + {\Lambda}$) one gets ${\beta}_{V^0} = 0.196 \pm 0.049$ which
significantly exceeds that for ${\pi}^-$ mesons. The ratio $R(V^0
/{\pi}^-) = <n_{V^0}> / <n_{{\pi}^-}>$, being equal to $0.087 \pm
0.015$ for the 'qusinucleon' interactions, increases up to $0.116
\pm 0.012$ (i.e. by a factor of about 1.3) at $A_{eff} \approx
21$, and up to $0.143 \pm 0.012$ (i.e. by a factor of about 1.6)
at $A_{eff} \approx 45$. Hence, the production of the neutral
strange particles is influenced by the nuclear medium stronger
than that for pions. A similar pattern was observed recently
\cite{ref7} in deep-inelastic neutrino-nucleus scattering (at $W
>$ 2 GeV and $Q^2 > $ 1 (GeV$/c)^2$). As it was shown in
\cite{ref7}, the $V^0$'s multiplicity gain ${\delta}_{V^0} =
{<n_{V^0}>}_A - {<n_{V^0}>}_N$ could be qualitatively explained in
the framework of a model incorporating the secondary intranuclear
interactions of produced pions, ${\pi}N \rightarrow V^0 X$, the
role of which turns out to be relatively more prominent, than that
for secondary interactions ${\pi}N \rightarrow {\pi}^- N$ which
results in a ${\pi}^-$ multiplicity gain ${\delta}_{\pi^-} =
{<n_{\pi^-}>}_A - {<n_{\pi^-}>}_N$ (see also \cite{ref8}). In the
next Section, a similar model predictions will be compared with
the multiplicity gains ${\delta}_{K^0}$, ${\delta}_{\Lambda}$ and
${\delta}_{\pi^-}$ extracted from the data of Table 1. \\ It is
expected, that the particles produced in secondary interactions
occupy predominantely the backward hemisphere in the hadronic
c.m.s. (i.e. the region of $y^* < 0$, $y^*$ being the particle
rapidity in that system). This expectation is verified by the data
on $<n_{V^{\circ}}>$ and $<n_{\pi^-}>$ for the both hemispheres
(Table 2).

\begin{table}[ht]
\begin{center}
\begin
{tabular}{|l| c c|}
  \hline
%\multicolumn{5}{|c|}{} \\
% &\multicolumn{4}{c|}{4 $< W^2 <$ 25
%GeV$^2$}\\  \multicolumn{5}{|c|}{} \\ $h^+(x_F > 0)$
$A_{eff}$&$<n_{V^{\circ}}>$&$<n_{\pi^-}>$ \\ \hline
&\multicolumn{2}{|c|}{$y^* > 0$}\\

 1&0.024$\pm$0.006&0.328$\pm$0.011 \\
21&0.030$\pm$0.005&0.308$\pm$0.007 \\ 45&0.042$\pm$0.005&$-$ \\
\hline

&\multicolumn{2}{|c|}{$y^* < 0$}\\

 1&0.024$\pm$0.005&0.224$\pm$0.009 \\
21&0.044$\pm$0.005&0.323$\pm$0.008 \\ 45&0.060$\pm$0.006&$-$ \\
\hline
\end{tabular}

\end{center}
\caption{The mean multiplicities $<n_{V^{\circ}}>$ and
$<n_{\pi^-}>$ at $y^* > 0$ and $y^* < 0$.}
\end{table}

\noindent In Fig. 2, the data of Table 2 are approximated by an
exponential dependence, resulting in ${\beta}_{V^0}(y^* > 0) =
0.147 \pm 0.074$, ${\beta}_{V^0}(y^* < 0) = 0.240 \pm 0.065$ and
${\beta}_{{\pi}^-}(y^* > 0) = -0.021 \pm 0.013$,
${\beta}_{{\pi}^-}(y^* < 0) = 0.120 \pm 0.016$. As it is seen from
Table 2 and Fig. 2, the nuclear effects induce a significant rise
of the $V^0$ multiplicity in the backward hemisphere and, to a
less extent, in the forward hemisphere. On the contrary, the
nuclear medium acts as an attenuator for the $\pi^-$ yield in the
forward hemisphere, while the A- dependence of $<n_{\pi^-}(y^* < 0
)>$ is significantly weaker as compared to $<n_{V^0}(y^* < 0)>$.
\\ A more relevant (conventional) measure of the nuclear
strangeness enhancement can be inferred from a comparison of the
relative yield  $R(K^{\circ}/\pi^-)$ =
$<n_{K^{\circ}}>/<n_{\pi^-}>$ in the 'quasinucleon' and nuclear
interactions. The data on $R(K^{\circ}/\pi^-)$, presented in Table
1 and Fig. 3, exhibit a noticeable A- dependence, which can be
described by a slope parameter ${\beta}_{K^0/{\pi}^-} = 0.157 \pm
0.070$. It is interesting to note, that in hadron-induced
reactions no nuclear enhancement of the ratio $R(K^0/\pi^-)$ was
observed \cite{ref9,ref10}, unlike to the charged kaon to pion
ratio which was found to be an increasing function of $A$
\cite{ref10,ref11,ref12}.
\\ More detailed information concerning nuclear effects in the
particle yield in different domains of the phase space can be
inferred from Figs. 4 - 7. Fig. 4 shows the rapidity
distributions. The distributions for the subsample $B_A$ are
shifted towards lower values of $y^*$ as compared to those for the
subsample $B_N$ (Figs. 4a - c). The nuclear enhancement effect is
more expressed at $y^* < -0.3$, being less significant at the
midrapidity ($|y^*| < 0.3)$. For the both ('quasinucleon' and
nuclear) subsamples, the ratio $R(K^{\circ}/\pi^-)$ tends to
increase with increasing $y^*$, being systematically higher for
the nuclear subsample. A faint indication is seen, that the
nuclear strangeness enhancement factor $R_A{(K^{\circ}/\pi^-)} /
R_N{(K^{\circ}/\pi^-)}$ is higher in the domain $y^* < -0.3$,
overlapping with the target fragmentation region. The
distributions on the kinematical variable $z = E_h/ \nu$ ($E_h$
being the energy of $K^0$ or $\Lambda$) are plotted in Fig. 5. It
is seen from Fig. 5a, that the nuclear enhancement effects for the
$K^0$ yield are significant at the low $z$ region ($z < 0.2 \div
0.3$), while for the leading $K^0$'s ($z > 0.4$) there is a faint
indication on a nuclear attenuation. These effects can be also
seen from Fig. 6a, which shows the slope parameter ${\beta}_{K^0}$
for $K^0$ mesons acquiring $z > z_{min}$. With increasing
$z_{min}$, the nuclear enhancement regime (${\beta}_{K^0} > 0$)
tends to be transformed to the attenuation one (${\beta}_{K^0} <
0$). Note, that the nuclear attenuation effects for charged kaons
with $z > 0.2$ were observed recently at higher energies ($7 < \nu
< 23$ GeV), in the deep-inelastic scattering of positrons on
nuclei \cite{ref13}. The data on $\Lambda$ are less conclusive
(Figs. 5b and 6b). They, however, indicate, that the $\Lambda$
yield at $z < 0.2 $ is definitely higher for the heaviest target
($A_{eff} \approx 45$).
\\ Fig.7 shows the squared transverse momentum distributions for
$K^0$ and $\Lambda$ ($p_T^2$ being measured transverse to the
current direction). They approximation by an exponential form,
$\sim \exp(-bp_T^2)$, results in $b_N(K^0) = 4.0\pm 1.4$ and
$b_N(\Lambda) = 4.1\pm 1.2$ (GeV$/c)^{-2}$ for the 'qusinucleon'
subsample and $b_A(K^0) = 4.0\pm 0.8$ and $b_A(\Lambda) = 3.3\pm
0.6$ (GeV$/c)^{-2}$ for the nuclear subsample. These values are
consistent with those obtained for neutrino-freon interactions
\cite{ref2}, $b_{Freon}(K^0) = 4.7 \pm 0.6$ and
$b_{Freon}(\Lambda) = 3.9 \pm 0.9$ (GeV$/c)^{-2}$, as well as for
(anti)neutrino interactions with protons \cite{ref5,ref6,ref14}
and light nuclei ($A \leq$ 20)
\cite{ref15,ref16,ref17,ref18,ref19}. Due to the large
uncertainties in our measurements, no conclusion can be drawn
concerning the A- dependence of $b(K^0)$ and $b(\Lambda)$.
However, the data of Fig. 7 indicate, that the nuclear enhancement
of the $K^0$ and $\Lambda$ yields occurs at comparatively low
$p_T^2$, at $p_T^2 <$ 0.4 (GeV$/c)^2$ for $K^0$ and $p_T^2 <$ 0.2
(GeV$/c)^2$ for $\Lambda$. One can, therefore, conclude, that
$K^0$'s and $\Lambda$'s produced in secondary intranuclear
interactions acquire on an average smaller $p_T$'s as compared to
those for the 'directly' produced ones.

\section{Discussion}

As it is shown in the previous Section, the multiplicity rise for
hadrons in nuclear interactions occurs mainly in the backward
hemisphere of the hadronic c.m.s. or, alternatively, in the
low-$z$ region. This behaviour can be conditioned by secondary
intranuclear collisions of 'primary' hadrons (mainly pions)
originated from the neutrino-nucleon interaction. \\ Table 3 shows
the experimental values of the total multiplicity gains
$\delta^{exp}_{K^0}, \delta^{exp}_{\Lambda}$ and
$\delta^{exp}_{\pi^-}$ (integrated over the whole phase volume)
for two composite nuclear targets with $A_{eff} \approx$ 21 and
45.

\begin{table}[ht]
\begin{center}
\begin
{tabular}{|l|c|c|c|c|c|c|}
  \hline
%\multicolumn{5}{|c|}{} \\
% &\multicolumn{4}{c|}{4 $< W^2 <$ 25
%GeV$^2$}\\  \multicolumn{5}{|c|}{} \\ $h^+(x_F > 0)$
$A_{eff}$&$\delta^{exp}_{K^0}$&$\delta^{th}_{K^0}$&$\delta^{exp}_{\Lambda}$&
$\delta^{th}_{\Lambda}$&$\delta^{exp}_{\pi^-}$&$\delta^{th}_{\pi^-}$
\\ \hline
21&\small{0.014$\pm$0.006}&\small{0.009$\pm$0.002}&\small{0.012$\pm$0.003}
&\small{0.008$\pm$0.002}&\small{0.087$\pm$0.010}&\small{0.108$\pm$0.023}
\\ \hline
45&\small{0.041$\pm$0.011}&\small{0.012$\pm$0.002}&\small{0.013$\pm$0.004}
&\small{0.010$\pm$0.002}&\small{0.168$\pm$0.016}&\small{0.146$\pm$0.031}
\\ \hline

\end{tabular}

\end{center}
\caption{The A - dependence of the experimental and calculated
multiplicity gains for $K^0$, $\Lambda$ and $\pi^-$.}
\end{table}

\noindent Bellow an attempt is undertaken to obtain predictions
for these gains in the framework of a model \cite{ref7,ref8}
incorporating secondary interactions of 'primary' pions, ${\pi}N
\rightarrow K^0 X$, ${\pi}N \rightarrow {\Lambda} X$, ${\pi}N
\rightarrow {\pi^-} X$. The mean multiplicity ${\bar{n}}_h(p_h)$
of the hadron $h$ ($h \equiv K^0, \Lambda$ or $\pi^-$) in
inelastic $\pi {N}$ interactions is estimated (with an uncertainty
of about 10-15 \%) from the available experimental data
\cite{ref20}, in the pion momentum range from $p_{\pi} \sim$ 0.9-1
GeV$/c$ up to $p_{\pi} \sim$ 11-12 GeV$/c$ (above which the pion
yield is negligible in this experiment). The corresponding mean
multiplicity in the ${\pi}^0$- induced reactions is assumed to be
the average of those in $\pi^+$ and $\pi^-$- induced reactions,
with an uncertainty related to the difference between the
latters.\\ The probability $w_A(p_{\pi})$ of the secondary
inelastic interactions of pions within nuclei ($A\equiv C, F, Br$)
is calculated taking into account their formation length
\cite{ref21}. The obtained $A$- dependence of $w_A(p_{\pi})$ and
the mean probability $<w_A(p_{\pi})>$, averaged over the nuclei of
the propane-freon mixture or the freon, are used to estimate the
effective atomic weight $A_{eff}$ of the composite target from the
requirement that the probability of the inelastic interaction of
pions within the nucleus with $A=A_{eff}$ is equal to
$w_{A_{eff}}(p_{\pi}) = <w_A(p_{\pi})>$. As a result, one obtains
$A_{eff} = 21\pm 2$ and 45$\pm$2, respectively, for the
propane-freon mixture and the freon, the quoted errors reflecting
the fact that the extracted value of $A_{eff}$ turns out to be
slightly dependent on the pion momentum. \\ In order to obtain the
contribution from the secondary interactions of ${\pi}^{\pm}$
mesons to the multiplicity gain, ${\delta}_h({\pi}^{\pm} N)$, the
product $w_A(p_{\pi}) \cdot \bar{n}_h(p_{\pi})$ was integrated
over the momentum spectra of  'primary' $\pi^+$ and $\pi^-$ mesons
measured in the 'quasinucleon' interactions. The expected
contribution of non-identified protons (estimated with the help of
the LEPTO 6.5 event generator \cite{ref22}) was subtracted from
the $\pi^+$ spectrum. Besides, the absorption of low energy
$\pi^-$ mesons on the quasinucleon pairs within the nucleus was
taken into account (see for details \cite{ref23,ref24} and
references therein). The contribution from the secondary
interactions of the 'primary' $\pi^0$ mesons was estimated under
assumption that their momentum spectrum is the average of those
for $\pi^+$ and $\pi^-$ mesons (see \cite{ref25} and references
therein). \\ Table 3 shows the summary contribution from the
secondary interactions of $\pi^+$, $\pi^-$ and $\pi^0$ mesons to
the multiplicity gains ${\delta}_{K^0}^{th}$,
${\delta}_{\Lambda}^{th}$ and ${\delta}_{\pi^-}^{th}$, the latter
being reduced due to the negative contribution from the $\pi^-$
absorption (note, that about 8\% and 11\% of 'primary' $\pi^-$
mesons are estimated to be absorbed in nuclei of the propane-freon
mixture and the freon, respectively). \\ It is seen from Table 3,
that the predicted multiplicity gains are compatible with the
experimental values, except for ${\delta}_{K^0}^{th} = 0.012 \pm
0.002$ at $A_{eff} \approx 45$ which composes only 29$\pm$9\% of
the measured value ${\delta}_{K^0}^{exp} = 0.041 \pm 0.011$. In
Fig. 8, the predicted A- dependence of the total mean multiplicity
${<n_h>}_A^{th} = {<n_h>}_N + {\delta}_h^{th}$ is compared with
the experimental data, which also include those for $\nu Ne$-
interactions, extracted from \cite{ref15} (where the same cut $W
>$ 1.5 GeV was applied) via an approximation of the measurement
results around $E_{\nu} =$ 9.8 GeV (the mean neutrino energy in
our experiment). It is seen, that the model qualitatively
describes the A- dependence of the yields of $\Lambda$ and
$\pi^-$, but predicts lower values for the $K^0$ yield as compared
to the data. Unlike the experimentally observed nuclear
enhancement of the ratio $R(K^0/\pi^-)$ (the last column of Table
1), no A- dependence for this ratio results from the model; the
predicted values of $R^{th}(K^0/\pi^-) = 0.059 \pm 0.011$ and
0.059 $\pm$ 0.010 at $A_{eff} \approx 21$ and 45, respectively,
are consistent with $R_N(K^0/\pi^-) = 0.055 \pm 0.013$ (cf. Table
1). \\ One can, therefore, conclude, that the secondary
intranuclear interaction processes, incorporated in the applied
model, are far from to be sufficient to explain the observed
enhancement of the $K^0$ yield. It seems, that the others (not
considered in the model) processes cannot improve radically the
data description. For example, in the secondary processes like
$K^+ n \leftrightarrow K^0 p$, $K^- p \leftrightarrow \bar{K^0}
n$, $Y N \leftrightarrow \bar{K^0} N'$ (where $Y$ stands for a
hyperon), the contributions of the direct and inverse reactions
should largely cancel each other. The model does not also
incorporate the production of hadronic resonances with a proper
space-time structure of their formation, intranuclear interactions
and decay. However, even if a resonance, e.g. $\rho$ meson (which
yield composes about 10\% of that for charged pions
\cite{ref26,ref27}) produces more $K^0$'s in a reaction $\rho N
\rightarrow K^0 X$ as compared to  $\pi N \rightarrow K^0 X$, the
contribution from the latter turns out to be reduced due to the
reduced multiplicity of the 'primary' pions, hence resulting in a
partly compensation of the contribution from the former. As for
the intranuclear interactions of 'primary' recoil nucleons, $N N
\rightarrow K^0 X$, their contribution to ${\delta}^{th}_{K^0}$
has been estimated to be less than a few percents as compared to
that for 'primary' pions.
\\
A better description of the data on $\delta^{exp}_{K^0}$ can be
achieved under assumption that the quark string fragmentation is
influenced by the surrounding nuclear medium, the quark string
tension $\kappa$ being dependent on the local nuclear density
${\rho}_A(r)$ near the string breaking point. A simplest
assumption is a linear dependence ${\kappa}_A(r) = {\kappa}_N[1+a
\, {\rho}(r)]$, where ${\kappa}_N \approx 1~GeV/fm$ is the string
tension in the vacuum, while $a$ is a proportionality coefficient.
Below we will present the model predictions at $a = 1/(0.6~fm^3)$.
At this value, the string tension ${\kappa}_A$ for the case of C,
F and Br nuclei turns out to be ${\kappa}_C = 1.158$, ${\kappa}_F
= 1.180$ and ${\kappa}_{Br} = 1.198$. In their turn, the latters
correspond to a nuclear enhancement of the $(s \bar{s})$ - pair
yield characterized by the ratio $r_{\lambda}(A) =
{\lambda}_A^{eff}/{\lambda}_N$, where ${\lambda}_A^{eff}$ is
related to the Wroblewsky parameter ${\lambda}_N$ (relevant for
the string fragmentation in the vacuum) \cite{ref28} as
${\lambda}_A^{eff} = {\lambda}_N^{{\kappa}_N/{\kappa}_A}$
\cite{ref29}. Taking into account a possible uncertainty in
${\lambda}_N$, $0.15 < {\lambda}_N <0.20$, the ratio
$r_{\lambda}(A)$ for the case of C, F and Br nuclei turn out to be
$1.25 < r_{\lambda}(C) < 1.30$, $1.28 < r_{\lambda}(F) < 1.34$ and
$1.31 < r_{\lambda}(Br) < 1.37$. The corresponding multiplicity
rise for $V^0$'s as compared to the mean multiplicity
$<n_{V^0}>_N$ in neutrino interactions with (quasi)free nucleons
is, therefore, $(r_{\lambda}(A) - 1) \cdot {<n_{V^0}>}_N$. The
latter is added to the contribution from the intranuclear
secondary interactions (given in Table 3). The resulting predicted
values of ${\delta}_h^{th}$ are presented in Table 4. It is seen,
that the model predictions at the increased string tension lead to
a comparatively better description of the data on the whole.
However, the predicted value for ${\delta}_{K^0}$ at $A_{eff}
\approx 45$, ${\delta}_{K^0}^{th} = 0.022 \pm 0.004$, is still
significantly underestimated, composing only $54 \pm 18$\% of the
measured value. \\

\begin{table}[ht]
\begin{center}
\begin{tabular}{|l|c|c|c|c|c|c|}
  \hline
%\multicolumn{5}{|c|}{} \\
% &\multicolumn{4}{c|}{4 $< W^2 <$ 25
%GeV$^2$}\\  \multicolumn{5}{|c|}{} \\ $h^+(x_F > 0)$
 { $A_{eff}$}&{ $\delta^{exp}_{K^0}$}&{
$\delta^{th}_{K^0}$}&{ $\delta^{exp}_{\Lambda}$}& {
$\delta^{th}_{\Lambda}$}&{$\delta^{exp}_{\pi^-}$}&{
$\delta^{th}_{\pi^-}$}
\\ \hline
 21&{\small 0.014$\pm$0.006}&{\small 0.018$\pm$0.003}&{\small0.012$\pm$
 0.003} &{\small
0.013$\pm$0.003}&{\small 0.087$\pm$0.010}&{\small 0.112$\pm$0.024}
\\ \hline
45&\small{0.041$\pm$0.011}&\small 0.022$\pm$0.004& \small
0.013$\pm$0.004& \small 0.016$\pm$0.003&\small 0.168$\pm$0.016&
\small 0.151$\pm$0.032 \\ \hline

\end{tabular}

\end{center}

\caption{The A - dependence of the experimental and calculated (at
an increased string tension, see text) multiplicity gains for
$K^0$, $\Lambda$ and $\pi^-$.}
\end{table}

\noindent It should be also noted, that a small difference between
the values of ${\delta}_{\pi^-}^{th}$ given in Tables 3 and 4 is
caused by a slightly different probabilities of the secondary
intranuclear interactions due to the difference in the formation
length of pions (being inversely proportional to the string
tension $\kappa$ \cite{ref21}).

\section{Summary}

For the first time, the A- dependence of the neutral strange
particle and, for comparison, $\pi^-$ meson production in the
neutrinonuclear interactions is investigated. This dependence for
$K^0$ mesons and $\Lambda$ hyperons is found to be significantly
stronger than for $\pi^-$ mesons. An exponential approximation
($\sim A^{\beta}$) of the particle yields results in a slop
parameter $\beta_{V^0} = 0.196 \pm 0.049$ for the summary yield of
$K^0$ and $\Lambda$ in the whole phase volume, while for $\pi^-$
mesons this parameter is noticeable smaller, $\beta_{\pi^-} =
0.068 \pm 0.007$. A nuclear enhancement of the $K^0/{\pi^-}$ ratio
is observed; this ratio, being equal to $R_N(K^0/{\pi^-}) =
0.055\pm 0.013$ in $\nu N$- interactions, increases with the
target atomic weight, reaching $R_A(K^0/{\pi^-}) = 0.070\pm 0.011$
at $A \approx 21$ and 0.099$\pm$0.011 at $A \approx 45$. \\ It is
observed, that the multiplicity rise of the neutral strange
particles occurs predominantly in the backward hemisphere (in the
hadronic c.m.s.) or, alternatively, in the low $z$ region ($z <
0.2$), hence indicating on a prominent role of the secondary
intranuclear processes. An attempt is undertaken to estimate the
nuclear enhancement of the particle yields caused by the secondary
intranuclear collisions of 'primary' pions originated from
neutrino-nucleon interactions. The model calculations reproduce
the measured multiplicity rise for $\pi^-$ mesons and $\Lambda$
hyperons, but badly underestimate that for $K^0$ mesons. Somewhat
better description of the data can be achieved under assumption
that the process of the quark string fragmentation is influenced
by the surrounding nuclear medium, which induces an increasing of
the string tension and, hence, an enhancement of the $(s \bar{s})$
- pair yield. \\

\noindent {\bf{Acknowledgement:}} The activity of two of the
authors (H.G. and Zh.K.) is partly supported by Cooperation
Agreement between DESY and YerPhI signed on December 6, 2002. The
autors from YerPhI acknowledge the supporting grants of Calouste
Gulbenkian Foundation and Swiss Fonds "Kidagan".

%%%%%%%%%%%%%%%%%%%%%%%%%%%%%%%%%%%%%%%%%%%%%%%%%%%%%%%%%%
   %%% References
%%%%%%%%%%%%%%%%%%%%%%%%%%%%%%%%%%%%%%%%%%%%%%%%%%%%%%%%%%

\newpage

\begin{figure}
\resizebox{1.1\textwidth}{!}{\includegraphics*[bb=50 120 600
530]{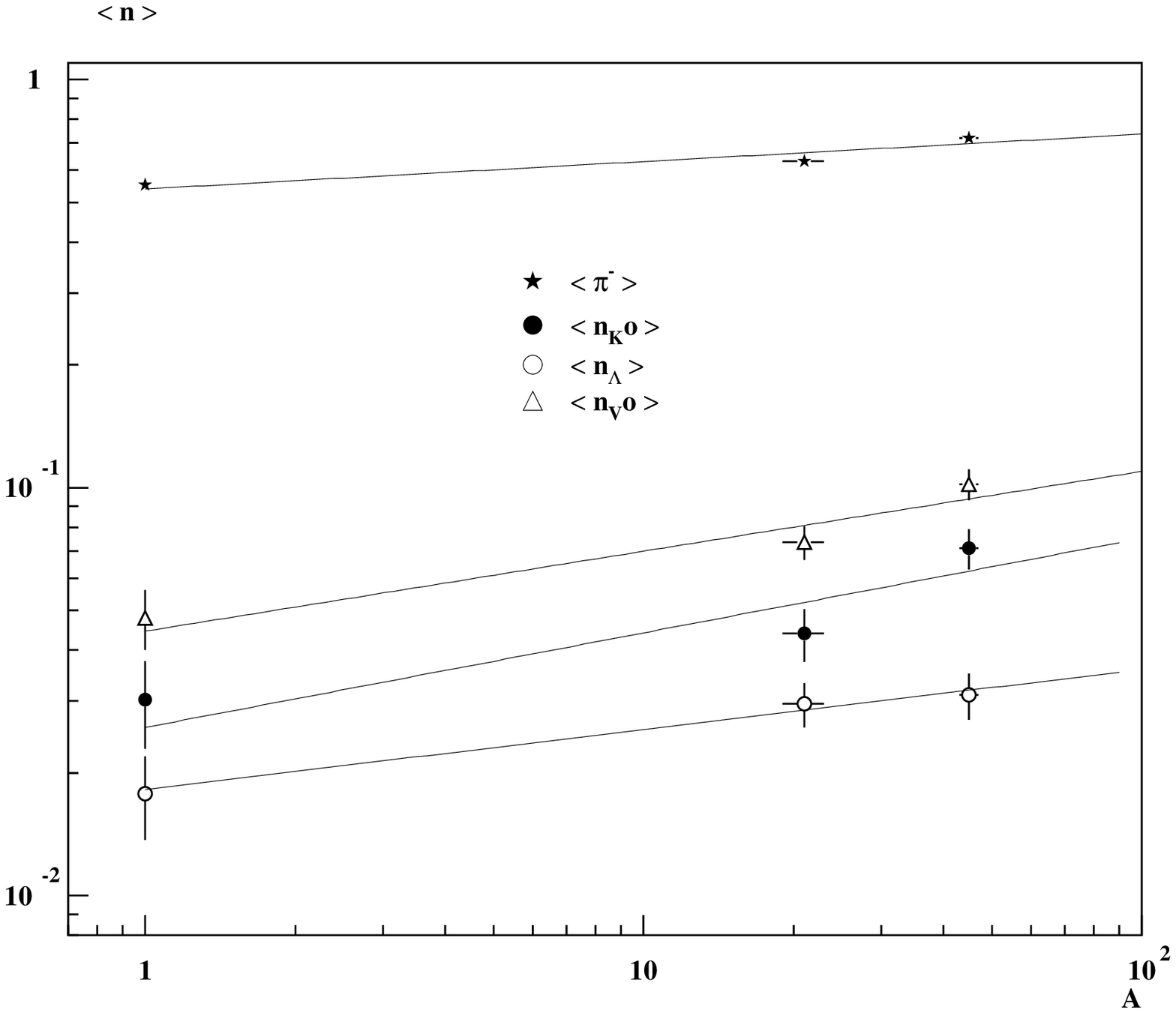}}  \vspace{0.5cm} \caption{The A- dependence of the
total yields of $K^0$, $\Lambda$, $V^0$ and $\pi^-$. The curves
are the result of the exponential fit.}

%\label{gmfig14}
\end{figure}

\newpage
\begin{figure}
\resizebox{1.1\textwidth}{!}{\includegraphics*[bb=50 120 600
530]{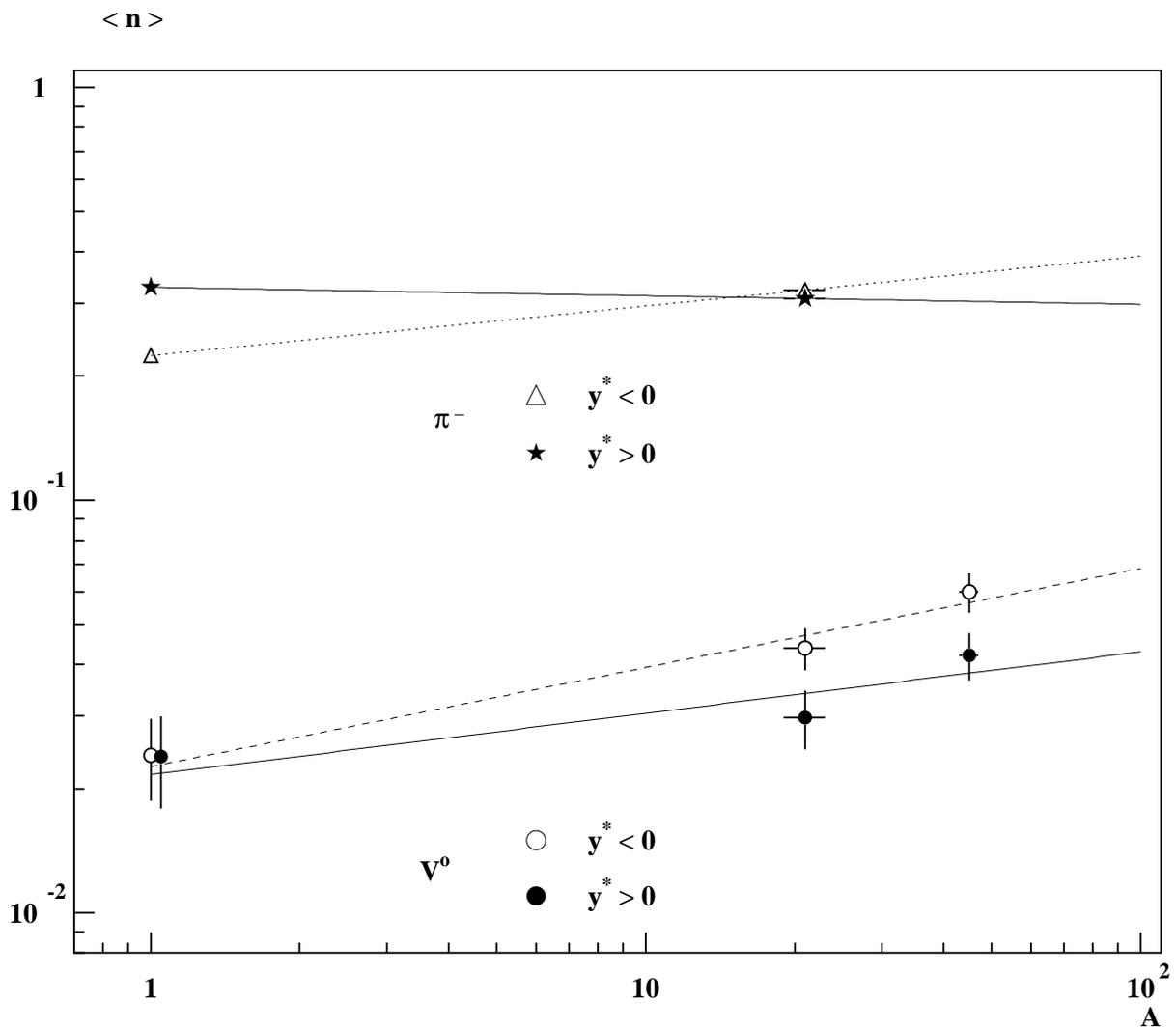}}  \vspace{0.5cm} \caption{ The A- dependence of the
yields of $V^0$, and $\pi^-$ in the forward ($y^* > 0$) and
backward ($y^* < 0$) hemispheres. The curves are the result of the
exponential fit.}
\end{figure}

\newpage

\begin{figure}
\resizebox{1.1\textwidth}{!}{\includegraphics*[bb=50 120 600
530]{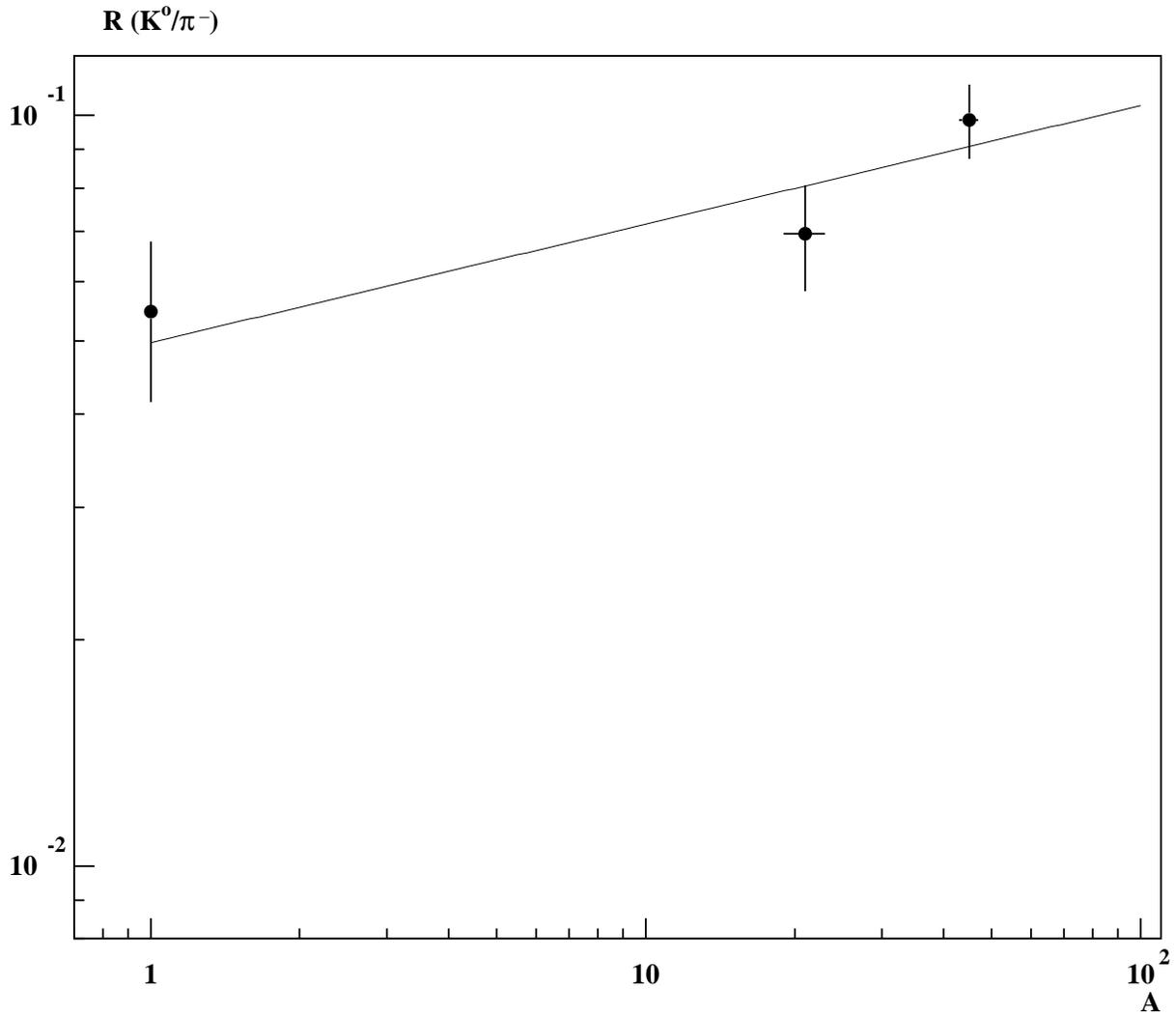}}  \vspace{0.5cm} \caption{The A- dependence of the
ratio $R(K^0/\pi^-)$. The curve is the result of the exponential
fit.}
\end{figure}

\newpage

\begin{figure}
\resizebox{1.1\textwidth}{!}{\includegraphics*[bb=50 120 600
530]{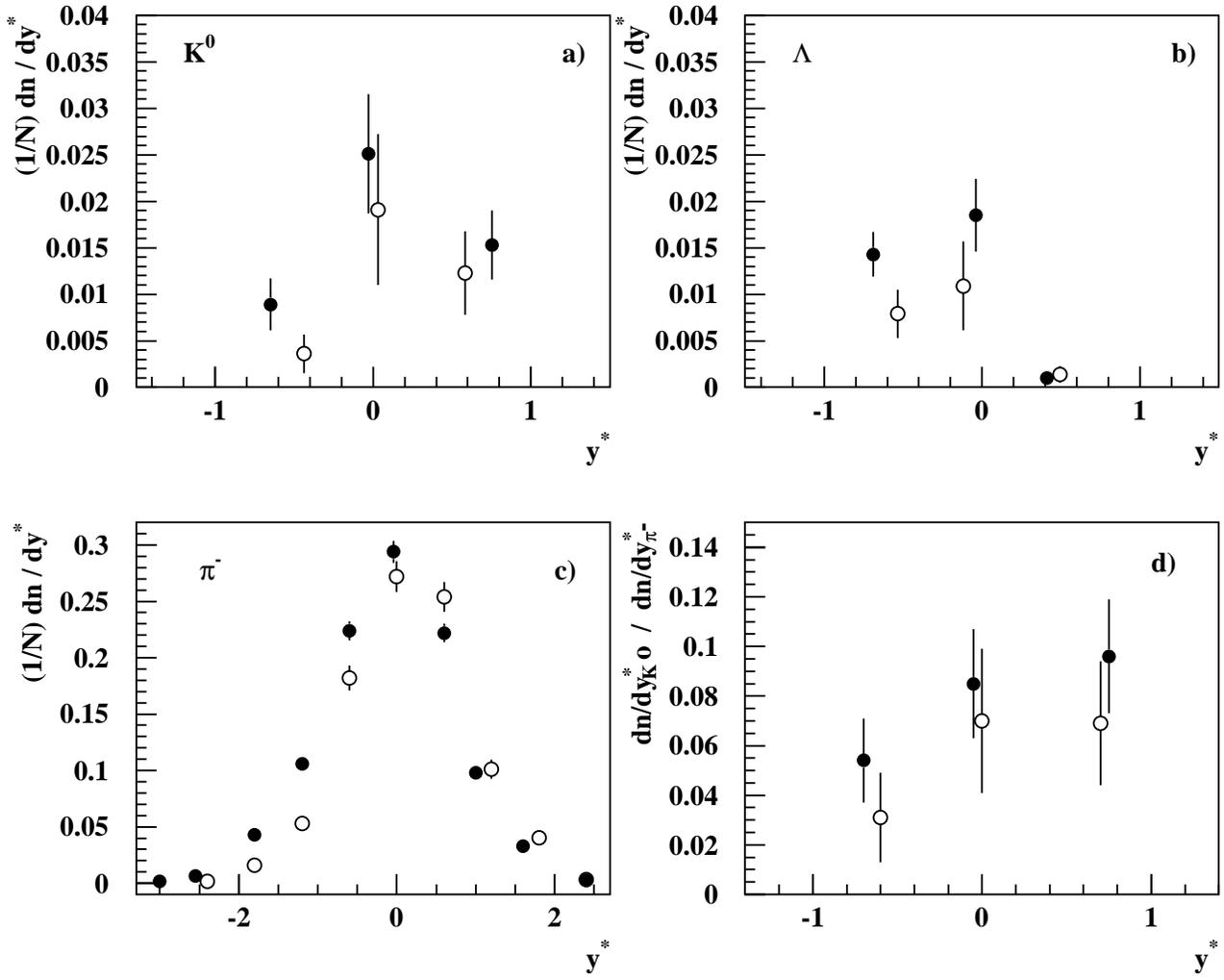}}  \vspace{0.5cm} \caption{ The rapidity
distributions of $K^0$ (a), $\Lambda$ (b), $\pi^-$ (c) and the
ratio $K^0/\pi^-$ (d) in the 'quasinucleon' (open circles) and
nuclear (black circles) subsamples.}
\end{figure}

\newpage

\begin{figure}
\resizebox{1.1\textwidth}{!}{\includegraphics*[bb=50 120 600
530]{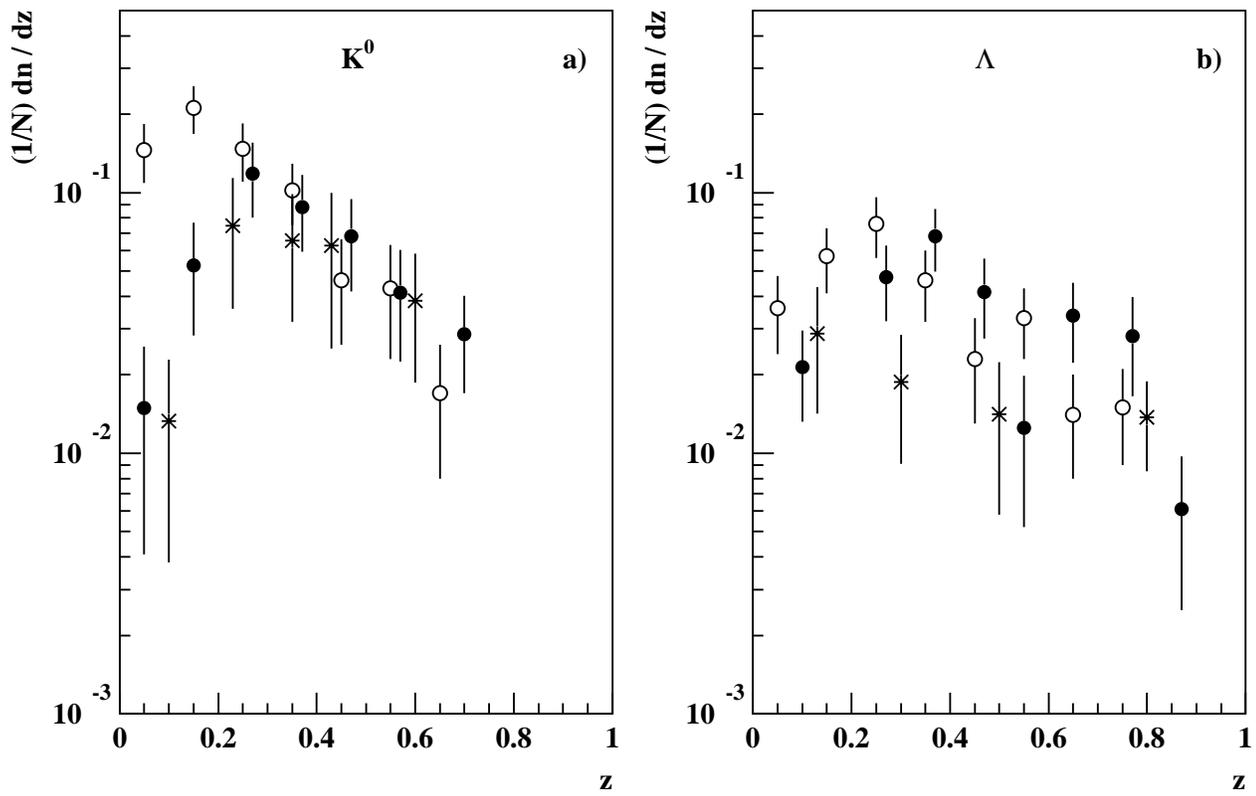}}  \vspace{0.5cm} \caption{The $z$ distributions of
$K^0$ (a) and $\Lambda$ (b) in the 'quasinucleon' subsample
(asterisks), at $A_{eff} \approx 21$ (black circles) and $A_{eff}
\approx 45$ (open circles).}
\end{figure}

\newpage

\begin{figure}
\resizebox{1.1\textwidth}{!}{\includegraphics*[bb=50 120 600
530]{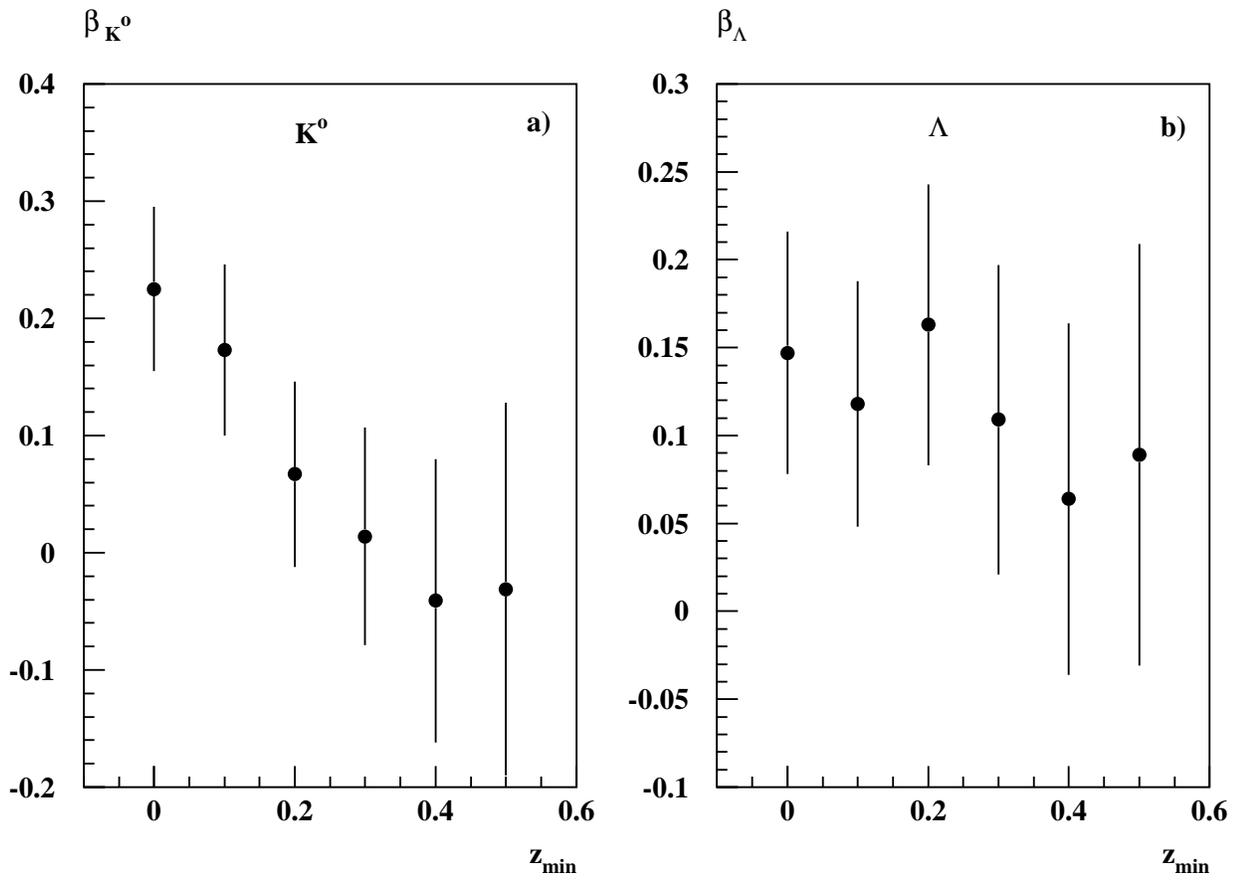}} \vspace{0.5cm} \caption{ The $z_{min}$- dependence
of the slope parameters ${\beta}_{K^0}$ (a) and
${\beta}_{\Lambda}$ (b).}
\end{figure}

\newpage

\begin{figure}
\resizebox{1.1\textwidth}{!}{\includegraphics*[bb=50 120 600
530]{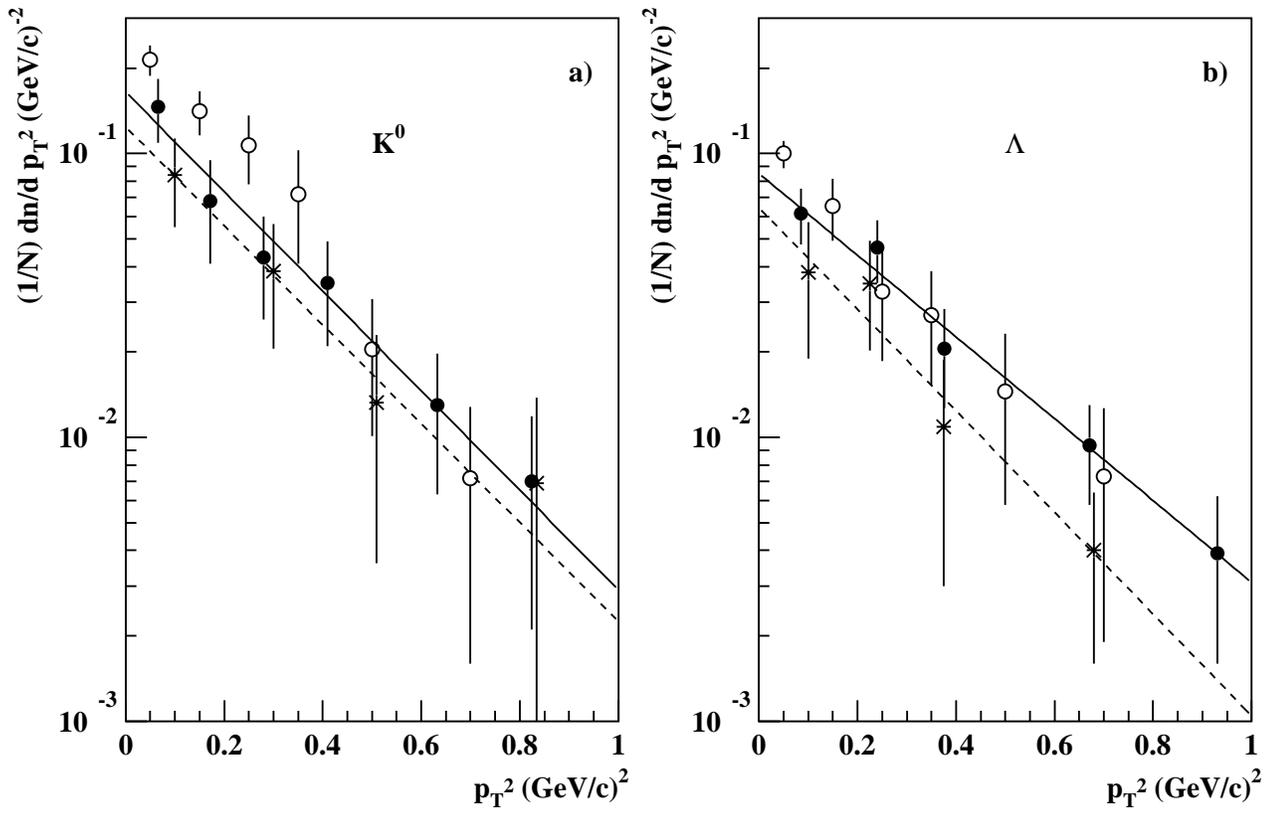}} \vspace{0.5cm} \caption{The $p_T^2$- distributions
of $K^0$ (a) and $\Lambda$ (b) in the 'quasinucleon' subsample
(asterisks), at $A_{eff} \approx 21$ (black circles) and $A_{eff}
\approx 45$ (open circles). The curves are the result of the
exponential fit for the 'quasinucleon' subsample (dashed) and for
$A_{eff} \approx 21$ (solid).}
\end{figure}

\newpage

\begin{figure}
\resizebox{1.1\textwidth}{!}{\includegraphics*[bb=50 120 600
530]{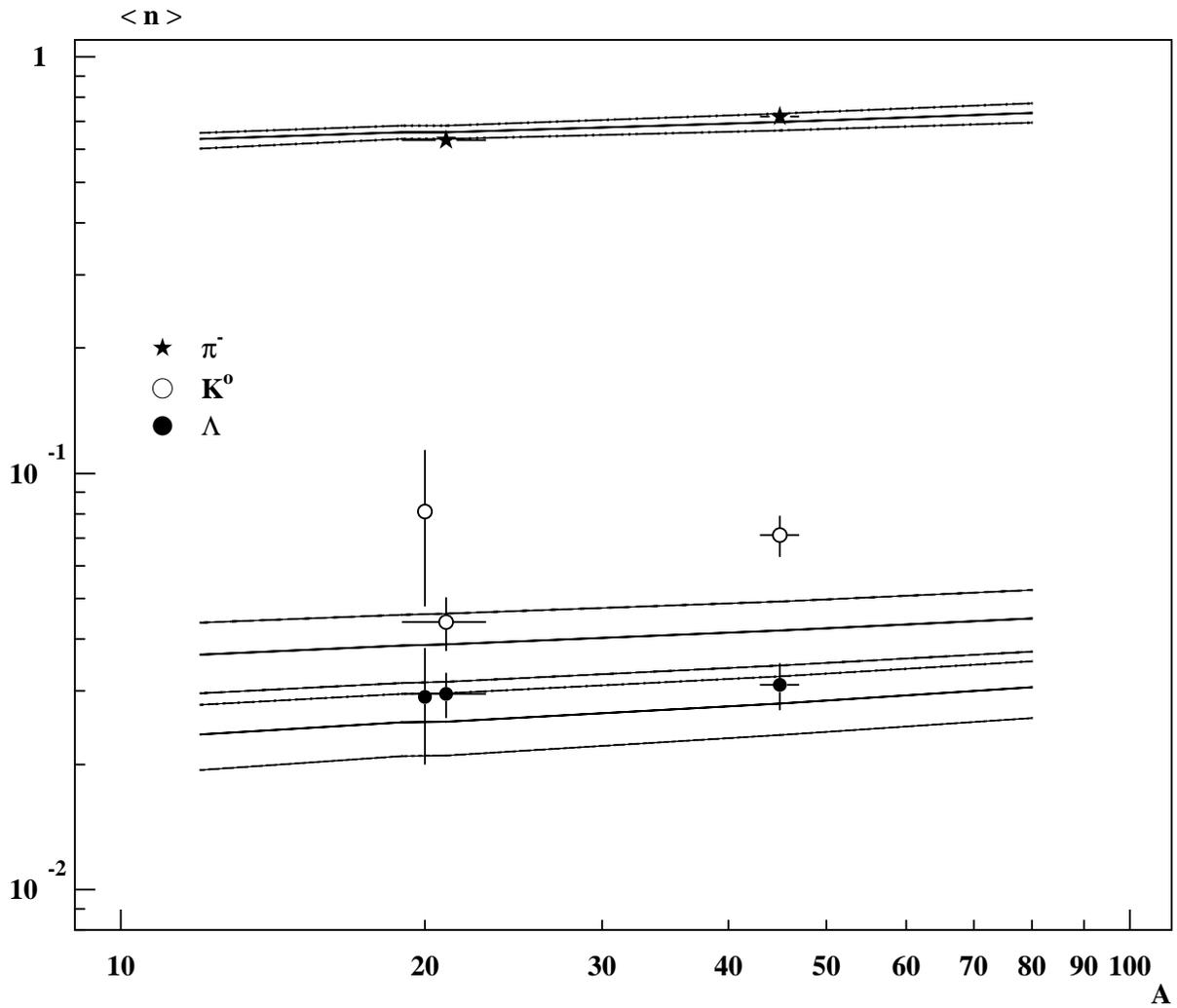}} \vspace{0.5cm} \caption{The A- dependence of the
total yields of $K^0$ (open circles), $\Lambda$ (black circles)
and $\pi^-$ (asterisks). The solid and dashed curves are the model
predictions and the uncertainty in the latter (see text).}
\end{figure}

\end{document}